\documentclass[10pt,a4paper]{article}

\usepackage[margin=2.5cm]{geometry}
\usepackage{times}
\usepackage{graphicx}
\usepackage{amsmath,amssymb}
\usepackage{authblk}
\usepackage[numbers,sort&compress]{natbib}
\usepackage{hyperref}
\title{\bfseries Exploring Quantum Corners:
How Curved Momentum Space Shapes BTZ Black
Holes}

\author[1,2]{Partha Nandi}

\affil[1]{\small Department of Physics, Stellenbosch University, South Africa}
\affil[2]{\small National Institute for Theoretical and Computational Sciences (NITheCS), South Africa}

\affil[ ]{\small E-mail: pnandi@sun.ac.za, partha.nandi@nithecs.ac.za}

\date{} 

\begin{document}

\maketitle

\begin{abstract}
Planck-scale quantum features of spacetime may admit effective semiclassical manifestations not only through spacetime geometry but also through the geometry of momentum space. In this work, we develop a tractable $(2+1)$-dimensional framework in which the ${\rm su}(1,1)$ algebra of spacetime localization operators provides an effective description of a particular quantum corner of the Planck-scale structure of spacetime. The semiclassical limit of this localization algebra naturally reconstructs a locally ${\rm AdS}_3$ momentum-space geometry. The resulting momentum-space curvature deforms the symplectic structure, modifies the particle dispersion relation, and gives rise to a finite renormalized mass. Using an effective configuration-space action, we derive the corresponding effective energy-momentum tensor and demonstrate that it consistently sources the classical Einstein equations. The resulting spacetime is a deformed BTZ black hole whose conserved ADM mass, horizon radius, Hawking temperature, and Bekenstein--Hawking entropy receive finite Planck-scale corrections encoded entirely in the nonlinear relation between the microscopic particle mass and the conserved gravitational mass. We further investigate Hawking emission using the Hamilton--Jacobi tunneling formalism and show that the return time of an emitted massless quantum receives two competing contributions: a geometric correction originating from curved momentum space and a dynamical correction arising from Hawking backreaction. An important outcome is that the null geodesic equations remain unchanged; the observable modifications originate entirely from the deformation of the effective spacetime geometry. Our results establish a concrete semiclassical mechanism through which the effective quantum kinematics encoded in curved momentum space is dynamically transferred into macroscopic gravitational phenomena through the unmodified Einstein equations, thereby providing a dynamical bridge between an effective manifestation of the quantum nature of spacetime and observable gravitational physics.
\end{abstract}


\section{Introduction}

General relativity describes gravity as the geometry of spacetime, whereas
quantum theory assigns a fundamental role to observables associated with
physical measurements. Reconciling these two principles remains one of the
central challenges in theoretical physics. A basic question is whether the
smooth spacetime manifold employed in classical gravity is itself
fundamental, or whether it emerges only as an effective description of a
more microscopic quantum structure. From an operational viewpoint \cite{PhysRevLett.47.979,Amelino-Camelia:2011lvm, Nandi:2026muw}, spacetime points are not directly
observable. Local observers measure quantities such as energies, momenta,
directions, arrival times, and detector correlations rather than spacetime
coordinates themselves. Coordinates acquire physical meaning only through
a reconstruction procedure based on these measurements together with
assumptions regarding locality, synchronization, and symmetry
\cite{AmelinoCamelia2011}. This observation naturally raises the possibility
that spacetime geometry may itself be emergent rather than fundamental \cite{Mohan:2026dah}.

One attractive possibility is that the primary geometric structure resides
not in spacetime but in momentum space. During the last decade, several
approaches to quantum gravity have suggested that momentum space may become
curved at the Planck scale
\cite{Kowalski-Glikman:2013rxa,Franchino-Vinas:2023rcc}. In particular,
within the framework of relative locality
\cite{Amelino-Camelia:2011lvm}, spacetime is regarded as a secondary
construction reconstructed from the cotangent bundle over momentum space.
In this picture, different observers need not agree on the localization of
distant events, and the geometry of momentum space directly influences the
effective notion of spacetime \cite{Nandi:2025qyj}. Similar ideas also realize Born's
reciprocity principle
\cite{RevModPhys.21.463}, placing spacetime and momentum space on an equal
conceptual footing.

The physical origin of curved momentum space has been discussed from
different perspectives. In the relative-locality programme and related
approaches, curved momentum space emerges as an effective description after
integrating out microscopic gravitational degrees of freedom, whose precise
nature depends on the underlying quantum-gravity theory. These microscopic
degrees of freedom may originate from spin-network variables, string
degrees of freedom, or other fundamental structures that remain unknown \cite{Scholtz2025, Eppley:1977emg}.
Regardless of their detailed realization, the resulting low-energy
description is naturally expressed in terms of a curved momentum-space
geometry.

A complementary viewpoint arises from the operational limitations of event
localization. Attempting to localize an event within a region of size
$\Delta x$ requires probes with momentum uncertainty
$\Delta p\sim\hbar/\Delta x$. Since the probe energy itself gravitates,
localization becomes increasingly affected by gravitational backreaction.
Combining the uncertainty principle with classical gravitational collapse
arguments leads to the Planck-scale bound
$\Delta x\gtrsim\ell_P$ \cite{Salecker:1958}. Below this scale,
localization procedures cannot be made arbitrarily sharp, suggesting that
the classical notion of spacetime points loses operational meaning \cite{PhysRev.135.B849, Townsend:1977xw}.

This observation was formalized by Doplicher, Fredenhagen, and Roberts
\cite{Doplicher:1994zv}, who argued that Planck-scale localization should
be described by noncommuting localization operators satisfying
\begin{equation}
[\hat x^a,\hat x^b]
=
i\theta^{ab}(\hat x).
\end{equation}
These operators should not be interpreted as coordinates of a fundamental
noncommutative spacetime. Rather, they encode the intrinsic quantum
limitations of localization itself \cite{PhysRevD.75.125020}. In this interpretation,
noncommutativity reflects the impossibility of simultaneously performing
arbitrarily precise localization experiments once gravitational
backreaction becomes significant
\cite{Nandi:2023tfq}.

Once localization outcomes cease to commute, the point-set description of
spacetime loses its operational significance. Geometry, however, need not
disappear; rather, it should be reconstructed from more fundamental
algebraic structures that remain well defined beyond the classical
localization limit. Related viewpoints have appeared in several different
approaches to quantum gravity. For example, in the connection formulation
of canonical quantum gravity, Ashtekar \textit{et al.} \cite{Ashtekar:1989qd, PhysRevLett.57.2244} regarded the
spacetime metric as a derived, secondary object reconstructed from more
fundamental gauge-theoretic variables rather than treated as the primary
quantity to be quantized directly. A mathematically
rigorous realization of this general philosophy is provided by Connes'
noncommutative geometry
\cite{Connes:1990qp,Connes:1994yd,Chakraborty:2021fbr,Chamseddine:2025wgr},
where geometric information is encoded in an involutive algebra, a Hilbert
space, and a Dirac operator rather than in a smooth point-set manifold. For semiclassical gravitational applications, however, a less elaborate
construction is often sufficient. Rather than reconstructing geometry
through the full machinery of spectral triples, one may start from a
noncommutative localization algebra, take its classical limit, and
reconstruct the effective geometric structures governing physical
observables \cite{Nandi:2025mco}.


An important realization of this philosophy was provided by Freidel and
Livine \cite{Freidel:2005bb,PhysRevLett.96.221301}, who showed that within the
Ponzano--Regge spin-foam model of $(2+1)$-dimensional quantum gravity \cite{PhysRevD.48.2702},
integrating over microscopic gravitational degrees of freedom naturally
produces an effective noncommutative field theory with a curved
(group-valued) momentum space. Their work demonstrates that curved momentum
space can emerge as the low-energy manifestation of a specific microscopic
quantum gravity model. However, despite many candidate theories, the
microscopic structure of quantum spacetime remains unknown. It is therefore
natural to investigate whether curved momentum space itself captures universal
low-energy signatures of Planck-scale physics without committing to a
particular microscopic realization.

The present work adopts this complementary effective-field-theory
perspective \cite{Wang:2023avy}. Our objective is not to derive curved momentum space from an
underlying microscopic quantum gravity model, but to investigate its physical
consequences once regarded as an effective manifestation of the unresolved
Planck-scale structure of spacetime. Starting from a noncommutative
$\mathfrak{su}(1,1)$ algebra describing localization observables, we show
that its classical limit reconstructs a locally AdS$_3$ momentum-space
geometry rather than a curved spacetime geometry
\cite{Nandi:2023deb}. The resulting momentum-space geometry induces
deformations of the symplectic structure and particle dispersion relations,
which can be consistently encoded within an effective configuration-space
action.

The principal aim of this paper is to establish the missing dynamical
connection between curved momentum space and gravity. While previous studies
of curved momentum space, doubly special relativity, and relative locality
have primarily focused on modified particle kinematics, localization, and
dispersion relations, we demonstrate how these effective quantum-kinematical
structures generate an effective stress-energy tensor that consistently
couples to the classical Einstein equations. The Einstein equations
themselves remain unchanged; all modifications arise from the effective
matter sector induced by the underlying momentum-space geometry. Coupling
this effective matter sector to $(2+1)$-dimensional Einstein gravity with a
negative cosmological constant \cite{Witten:1988hc}, we derive a BTZ
black-hole ~\cite{PhysRevLett.69.1849, Banados:1992gq} solution whose ADM mass, horizon radius, Hawking temperature, and
related thermodynamic observables receive finite Planck-scale corrections.
Our results therefore provide a complete semiclassical framework in which
curved momentum space acts as the intermediary between microscopic quantum
kinematics and observable gravitational phenomena.

The paper is organized as follows. In Sec.~II we construct the classical
Poisson limit of the noncommutative localization algebra and reconstruct the
emergent AdS$_3$ momentum-space geometry. Section~III derives the deformed
dispersion relation and the corresponding effective particle dynamics.
Section~IV develops the effective configuration-space action and the
associated stress-energy tensor. In Sec.~V we couple this matter sector to
$(2+1)$-dimensional Einstein gravity, derive the deformed BTZ solution, and
investigate its thermodynamic properties together with the propagation of
massless signals. Finally, Sec.~VI summarizes our results and discusses their
broader implications.


\section{Emergent Curved Momentum Space with Minimal Deformation}

We begin with a noncommutative algebra of localization operators,
denoted $\mathbb{R}^{(1,2)}_\star$, whose generators satisfy the
$\mathfrak{su}(1,1)$ algebra

\begin{equation}
[\hat{x}^a, \hat{x}^b] = i\,\epsilon^{ab}{}_{c}\hat{x}^c ,
\label{eq:su11}
\end{equation}
with signature $(-,+,+)$ and convention $\epsilon^{0}{}_{12}=1$.
This algebra is stable under Lorentz transformations
$\mathcal{SO}(1,2)$, $\hat{x}'^{\,a}=\Lambda^{a}{}_{b}\hat{x}^b$.

At this stage no spacetime manifold or metric geometry is assumed.
The operators $\hat{x}^a$ are interpreted as localization operators, and
the algebra encodes only their symmetry properties rather than geometric
data. In $(2+1)$ dimensions, $\mathfrak{su}(1,1)$ provides the minimal
noncompact Lie--algebraic structure whose automorphism group is
$\mathrm{SO}(1,2)$, ensuring Lorentz--covariant localization in a purely
algebraic sense. Among three--dimensional Lie algebras,
$\mathfrak{su}(1,1)$ is the simplest choice admitting a controlled
classical (Poisson) limit and noncompact coadjoint orbits, which are
essential for describing unbounded localization and momenta.
Compact algebras such as $\mathfrak{su}(2)$ would instead lead to bounded
localization operators and a compact momentum space, incompatible with
relativistic kinematics. Geometry is therefore not postulated at this
stage, but reconstructed only at a later semiclassical level.

To identify translation generators compatible with
\eqref{eq:su11}, we introduce commuting translation generators
\begin{equation}
[\hat{p}_a,\hat{p}_b]=0,
\label{eq:pp}
\end{equation}
and impose all Jacobi identities involving
$(\hat{x},\hat{x},\hat{p})$ and $(\hat{p},\hat{p},\hat{x})$.
Constant translations $\delta_\xi\hat{x}^a=\xi^a$ cannot be generated if one insists on
preserving the primitive noncommutative Lie-algebraic structure of the
spacetime coordinate algebra
unless this is momentum dependent. A general ansatz
\[
\delta_\xi\hat{x}^a=i[\xi^b \hat{p}_b , \hat{x}^a]
= \xi^a + \alpha(\xi\!\cdot\!\hat{p})\hat{p}^a
+ \beta\,\epsilon^{abc}\hat{p}_c\xi_b
\]
yields, upon enforcing the Jacobi identities,
\begin{equation}
\alpha=-\frac14,\qquad \beta=\frac12.
\end{equation}
The resulting deformed Heisenberg algebra is
\begin{equation}
[\hat{x}^a,\hat{p}_b]=i\,(E^{-1}(p))^{a}{}_{b},
\qquad
[\hat{p}_a,\hat{p}_b]=0,
\label{eq:xp}
\end{equation}
with
\begin{equation}
(E^{-1}(p))^{a}{}_{b}
= \delta^{a}{}_{b}
-\frac14\,\hat{p}^{a}\hat{p}_{b}
+\frac12\,\epsilon^{a}{}_{bc}\hat{p}^{c}.
\label{eq:Einv}
\end{equation}

It is worth emphasizing that infinitesimal translations are implemented as inner derivations of the
full phase-space algebra $\mathcal{A}=\langle\hat{x}^a,\hat{p}_b\rangle$,
\begin{equation}
\delta_\xi A := i[\xi^a \hat{p}_a , A], \qquad A\in\mathcal{A}.
\end{equation}

Although these transformations do not act as automorphisms of the
coordinate subalgebra generated by $\hat{x}^a$ alone, they are
well-defined derivations of the full phase-space algebra
$\mathcal{A}=\langle\hat{x}^a,\hat{p}_b\rangle$.
The commutativity of the momentum generators,
Eq.~\eqref{eq:pp}, implies that translations commute as derivations,
\begin{equation}
[\delta_\xi,\delta_\eta]=0 ,
\end{equation}
and therefore integrate to a well-defined Abelian group action
$T(\xi)=e^{\delta_\xi}$ with composition law
$T(\xi)T(\eta)=T(\xi+\eta)$.
This commutativity ensures homogeneity, in the sense that finite
translations are path independent and leave no residual dependence on
the order in which they are applied.
It thus provides an algebraic statement of the absence of any fixed
background structure: translations relate physically equivalent points
without introducing preferred locations or directions
\cite{PhysRevD.48.5721}.
Requiring the coordinate Lie algebra \eqref{eq:su11} to remain stable
and preserved in form under translations implies that translations
cannot act as automorphisms of the coordinate subalgebra.
This necessity enforces a deformation of the translation action and,
consequently, induces a momentum dependence in the mixed commutator
$[\hat{x}^a,\hat{p}_b]$.
This feature reflects a nontrivial geometric structure in momentum
space, as will become explicit in the following sections, rather than a
breakdown of translation symmetry.

Introducing a physical length scale via
$\hat{x}^a\rightarrow \hat{x}^a/(2\lambda)$ and
$\hat{p}^a\rightarrow (2\lambda/\hbar)\hat{p}^a$, the algebra becomes
\begin{align}
[\hat{x}^a,\hat{x}^b] &= 2i\lambda\,\epsilon^{ab}{}_{c}\hat{x}^c, \\
[\hat{p}_a,\hat{p}_b] &= 0, \\
[\hat{x}^a,\hat{p}_b] &= i\hbar\,(E^{-1}(p))^{a}{}_{b},
\end{align}
with deformation scale $1/m_p = 2\lambda/\hbar$.
In $(2+1)$ dimensions one has $m_p\sim 1/G$.

Taking the classical limit
\[
\{A,B\}=\lim_{\hbar\to 0}\frac{1}{i\hbar}[\hat{A},\hat{B}],
\]
we obtain the Poisson algebra on phase space
\begin{align}
\{x_a,x_b\} &= \frac{1}{m_p}\,\epsilon_{ab}{}^{c}x_c, \\
\{p_a,p_b\} &= 0, \\
\{x_a,p_b\} &= (E^{-1}(p))_{ab},
\label{eq:PB}
\end{align}
where
\begin{equation}
(E^{-1})_{ab}
= \eta_{ab}
+ \frac{1}{m_p}\,\epsilon_{ab}{}^{c}p_c
- \frac{1}{m_p^2}\,p_a p_b .
\label{eq:einvPB}
\end{equation}
The undeformed limit is recovered as $m_p\to\infty$.

At this stage, the algebraic structure is defined on full phase space
with coordinates $(x_a,p_\mu)$. However, phase space is not the manifold
on which we aim to introduce a metric geometry. Our goal is instead to
extract an effective geometry on \emph{momentum space}, which is a
manifold in its own right and not a submanifold of phase space.

Accordingly, we restrict attention to the subalgebra of phase-space
functions depending only on the momenta, $F=F(p)$. This restriction is
consistent, since the Poisson brackets \eqref{eq:PB} imply that the
action of the generators $x_a$ on momentum-dependent functions closes
within this subalgebra,
\begin{equation}
\{x_a,F(p)\}=(E^{-1})_{a}{}^{\mu}(p)\,\partial_\mu F(p).
\end{equation}
This provides a differential realization of the localization generators
directly from the inner derivations of the full phase-space algebra.
This perspective differs from the covariant-coordinate formulation of
Wess \textit{et al.} and the Poisson differential-calculus formulation
of Ho and Miao, where the differential structure is introduced in the
context of noncommutative gauge theory through covariant coordinates or
a compatible differential calculus, respectively~\cite{Madore:2000en,PhysRevD.64.126002}.

At this stage, it is convenient to make explicit our index conventions.
Greek indices $\mu,\nu,\ldots\in\{0,1,2\}$ are used to denote the
coordinates $p^\mu$ on momentum space $P_{m_p}$, which we treat as a
manifold in its own right. In contrast, Latin indices $a,b,\ldots$ label
components of vectors and tensors with respect to an orthonormal (but
generally non-holonomic) basis of the tangent space $T(P_{m_p})$ or its
dual cotangent space $T^*(P_{m_p})$. With this distinction,
$(E^{-1})_{a}{}^{\mu}(p)$ can be interpreted as a triad (the
$(2+1)$-dimensional analogue of a vielbein), relating the orthonormal
frame $\{x_a\}$ to the holonomic coordinate basis
$\{\partial/\partial p^\mu\}$ of the tangent space,
\begin{equation}
T(P_{m_p})=\mathrm{Span}\{x_a\}
=\mathrm{Span}\!\left\{\frac{\partial}{\partial p^\mu}\right\}.
\end{equation}

A standard result of Poisson geometry states that on a Poisson manifold
$(M,\{\cdot,\cdot\})$, the map $g\mapsto\{f,g\}$ defines a derivation of
the algebra $C^\infty(M)$ for any function $f$. Since derivations of
$C^\infty(M)$ are in one-to-one correspondence with vector fields on
$M$, each function $f$ uniquely determines a vector field
$X_f$ via $X_f(\cdot)=\{f,\cdot\}$ \cite{Vaisman1994,CannasdaSilva2001}.
Moreover, the Jacobi identity implies that the map $f\mapsto X_f$ is a
Lie-algebra homomorphism, satisfying
\begin{equation}
[X_f,X_g]=X_{\{f,g\}} .
\end{equation}

Applying this construction to the generators $x_a$, and restricting to
momentum-dependent functions, we obtain vector fields on momentum space.
We denote these vector fields by $X_{x_a}$, with differential
realization
\begin{equation}
X_{x_a} := (E^{-1})_{a}{}^{\mu}(p)\,\frac{\partial}{\partial p^\mu}.
\label{eq:vectorfield}
\end{equation}

These vector fields are linearly independent and satisfy the same
$\mathfrak{su}(1,1)$ algebra as the Poisson generators
\eqref{eq:PB}, now realized as Lie brackets of vector fields on momentum
space.

Since curvature is a metric notion, we introduce a momentum-space metric
by declaring this non-holonomic frame to be orthonormal,
\begin{equation}
g(X_{x_a},X_{x_b})=\eta_{ab}.
\label{eq:orthonormal}
\end{equation}

This condition uniquely determines the coordinate components of the
metric $g_{\mu\nu}(p)=g(\partial_\mu,\partial_\nu)$,
\begin{equation}
g_{\mu\nu}(p)
=
\frac{m_p^2}{m_p^2-p^2}
\left(
\eta_{\mu\nu}
+
\frac{p_\mu p_\nu}{m_p^2-p^2}
\right),
\label{eq:fullmetric}
\end{equation}
with inverse
\begin{equation}
g^{\mu\nu}(p)
=
\left(1-\frac{p^2}{m_p^2}\right)
\left(
\eta^{\mu\nu}
-
\frac{p^\mu p^\nu}{m_p^2}
\right).
\label{eq:invmetric}
\end{equation}

Because the metric is constructed entirely from the frame $\{x_a\}$ and
their inner products are constant, it follows that
$\mathcal{L}_{X_{x_{a}}}g=0$. The vector fields $x_a$ therefore generate
isometries of momentum space and form a Killing algebra.

To elucidate the geometric character of this metric, we embed momentum
space into flat $(1+3)$-dimensional Minkowski space with coordinates
$p^M=(p^a,p^3)$ and constraint $G_{MN}p^Mp^N=m_p^2$. The induced metric
on this hypersurface is
\begin{equation}
\tilde{g}_{\mu\nu}(p)
=
\eta_{\mu\nu}
+
\frac{p_\mu p_\nu}{m_p^2-p^2},
\end{equation}
which is locally $dS_3$. The full metric \eqref{eq:fullmetric} is
conformal to this metric with conformal factor
\begin{equation}
\Omega(p)=\frac{1}{\sqrt{1-p^2/m_p^2}}.
\end{equation}
The Ricci tensor and scalar curvature are
\begin{equation}
R_{\mu\nu}=-\frac{2}{m_p^2}g_{\mu\nu},
\qquad
R=-\frac{6}{m_p^2},
\end{equation}
establishing that momentum space is locally $\mathrm{AdS}_3$ with
curvature radius $m_p$, becoming flat in the commutative limit
$m_p\to\infty$.

Finally, we note that alternative choices of function space would lead to
different geometric interpretations. Restricting to
coordinate-dependent functions would induce a momentum-dependent (and
hence generally Finsler-type) geometry on configuration space, while
allowing general phase-space functions would naturally lead to a
symplectic, rather than metric, geometry on phase space. In the present
work we deliberately focus on momentum-dependent functions in order to
isolate the geometry of momentum space and its minimal deformation.

\section{Particle Dynamics and Deformed Dispersion Relation}

Before introducing an effective configuration--space action, it is
important to clarify the status of the variables used to describe
particle dynamics and their relation to spacetime geometry. In the
present framework, phase space---rather than spacetime---is taken as the
fundamental arena. When momentum space is curved, spacetime localization
is not a primitive notion but an inferred, observer-- and
probe--dependent construct, in the spirit of relative locality.

The noncommutative variables $x^a$ satisfy the Lie--Poisson algebra
\begin{equation}
\{x^a, x^b\} = \frac{1}{m_p}\,\epsilon^{ab}{}_{c}\, x^{c},
\end{equation}
whose Poisson tensor has rank~2, leaving one null direction. The space
parametrized by $x^a$ therefore carries only a Poisson structure and is
foliated by two--dimensional coadjoint orbits; it does not admit a
nondegenerate Lorentzian metric on the full three--dimensional manifold.
For this reason, the variables $x^a$ cannot be interpreted as spacetime
coordinates suitable for defining local geometric objects such as an
energy--momentum tensor.

This motivates the introduction of momentum--dependent Darboux
(Bopp--shifted) variables
\begin{equation}
q^\mu := x^a E_a{}^{\ \mu}(p),
\qquad 
\{q^\mu,q^\nu\}=0,
\qquad
\{q^\mu,p_\nu\}=\delta^\mu_{\ \nu},
\end{equation}
which define local holonomic (commutative) coordinates on phase space.
The existence of such coordinates is guaranteed locally by Darboux's
theorem and does not imply the existence of a global,
observer--independent spacetime manifold. Their momentum dependence is a
direct reflection of the underlying curved momentum--space geometry and
is a characteristic feature of relative locality. Noncommutativity is
not lost in this description: it is encoded in the momentum--dependent
vielbein $E_a{}^{\ \mu}(p)$, the deformed symplectic structure, and the
resulting modified dispersion relation. The variables $q^\mu$ should
therefore be understood as effective, probe--dependent configuration--
space coordinates appropriate for a semiclassical description, rather
than as fundamental spacetime coordinates.

\medskip

We next construct the Lorentz generators
\begin{equation}\label{31}
M_{\mu\nu} := q_{\mu} p_{\nu} - q_{\nu} p_{\mu},
\end{equation}
which satisfy the $\mathfrak{iso}(1,2)$ algebra:
\begin{align}\label{32}
[ M^{\mu\nu}, M^{\rho\sigma} ] &=
 i \left( 
 \eta^{\mu\rho} M^{\nu\sigma} - \eta^{\mu\sigma} M^{\nu\rho}
 - \eta^{\nu\rho} M^{\mu\sigma} + \eta^{\nu\sigma} M^{\mu\rho}
 \right), \\
[P^\mu, M^{\nu\lambda}] &= 
 i\left( \eta^{\mu\nu} P^\lambda - \eta^{\mu\lambda} P^\nu \right).
\end{align}

There are two Casimir invariants: the Pauli--Lubanski scalar
$W=\frac{1}{2}\epsilon^{\mu\nu\rho}P_\mu M_{\nu\rho}$, which vanishes for a
spinless particle, and the invariant $p_\mu p^\mu$. For a spinless
particle, only
\begin{equation}
p_\mu p^\mu = -m^2
\end{equation}
is relevant.

When momentum space is curved, this invariant is deformed. Lorentz
covariance requires the deformed dispersion relation to take the form
$f(p^2)$, reducing smoothly to the commutative limit.

We therefore consider the following first--order Lagrangian, describing
the dynamics of a single spinless but massive point particle in an
effective configuration--space description:
\begin{equation}\label{26}
L = -q_{\mu}\, \dot{p}^{\mu} - \Lambda\!\left(f(p^2) + M^2\right), 
\qquad 
\dot{p}^{\mu} := \frac{dp^{\mu}}{d\tau},
\end{equation}
where $\tau$ is the worldline parameter and $\Lambda$ is a Lagrange
multiplier enforcing the constraint
\begin{equation}\label{27}
f(p^2) + M^{2} \approx 0,
\qquad 
M^2 = f(p^2),
\end{equation}
with $f$ an invertible function to be determined.

To determine $f(p^2)$, we identify it with the squared geodesic distance
in momentum space from the origin,
\begin{equation}
D = \sup_\gamma \int_0^P 
\sqrt{-g_{\mu\nu}(p')\, dp'^\mu dp'^\nu},
\end{equation}
which satisfies the Hamilton--Jacobi--type equation
\begin{equation}\label{34}
(\partial_{\mu} C)\, g^{\mu\nu}(p)\, (\partial_{\nu} C) = -4C,
\qquad 
C = D^2.
\end{equation}
In Riemann--normal coordinates $\pi^a$ based at the origin,
\begin{equation}\label{35}
C = -\eta_{ab}\, \pi^a \pi^b,
\end{equation}
with
\begin{equation}\label{36}
\pi^a = n^a \sqrt{-f(p^2)}, 
\qquad 
n^a n_a = -1.
\end{equation}
Identifying $C(p)=-f(p^2)$ and imposing the correct commutative limit yields
\begin{equation}\label{38}
f(p^2) = -m_{p}^2 
\left[
\tan^{-1}\!\left( \frac{\sqrt{-p^2}}{m_p} \right)
\right]^2 
=:\ -M^2 .
\end{equation}

The resulting renormalized mass
\begin{equation}
M = m_p\, \tan^{-1}\!\left( \frac{m}{m_p} \right)
\end{equation}
is bounded,
\begin{equation}
0 < M < \frac{\pi}{2} m_p,
\end{equation}
providing a natural ultraviolet saturation scale in $(2+1)$--dimensional
gravity arising from momentum-space curvature.

\section{Effective Configuration--Space Action and Stress Tensor}

Having established the deformed dispersion relation and the bounded
renormalized mass $M$, we now construct an effective
configuration--space description suitable for coupling to gravity.
While the curved momentum-space geometry developed in the previous
sections governs the microscopic particle kinematics, the
gravitational dynamics are naturally formulated in terms of a
spacetime stress--energy tensor. It is therefore necessary to
translate the modified phase-space dynamics into an effective
configuration-space action depending only on the spacetime
coordinates. This effective action provides the bridge between the
microscopic quantum kinematics encoded in the curved momentum space
and the effective matter sector that sources the classical Einstein
equations.

We begin with the first-order worldline action
\begin{equation}
S[q,p,\Lambda]
=
\int d\tau
\left(
p_\mu\dot q^\mu
+
\Lambda
\left[
\mathcal C(p)-M^2
\right]
\right),
\label{eq:firstorder}
\end{equation}
where $\Lambda$ is a Lagrange multiplier enforcing the modified
mass-shell condition and $\mathcal C(p)$ denotes the Casimir
associated with the curved momentum-space geometry. Varying the action
with respect to the momentum yields
\begin{equation}
\dot q^\mu
=-
\Lambda
\frac{\partial\mathcal C}{\partial p_\mu},
\end{equation}
which can be perturbatively inverted to express the momentum in terms
of the worldline velocities. Eliminating the momentum variables and
the auxiliary Lagrange multiplier $\Lambda$, and substituting the
resulting expressions back into the first-order action, yields an
effective configuration--space action depending solely on the
particle trajectory. Since the reduced Lagrangian is not homogeneous
of degree one in the velocities, the resulting worldline action is
not manifestly reparametrization invariant off shell. Consequently,
its Euler--Lagrange equations initially take a non-affine form.
However, the equations of motion imply that the non-affine
contribution vanishes on shell, so that the physical trajectories are
governed by the affine geodesics of the background spacetime. The
details of this construction are presented in the appendices of
Ref.~\cite{Nandi:2025mco}. The resulting effective action is
\begin{equation}
S_{\rm eff}[q]
=
\int d\tau
\left[
-\alpha
\sqrt{-\dot q^2}
-
\beta
(-\dot q^2)^{5/2}
+
\mathcal O
\left(
\frac{1}{m_p^4}
\right)
\right],
\label{Seff-flat}
\end{equation}
where
\begin{equation}
\alpha
=
M
\left(
1+\frac{M^2}{3m_p^2}
\right),
\qquad
\beta
=
\frac{M^3}{3m_p^2},
\end{equation}
and
\[
\dot q^2
=
\eta_{\mu\nu}
\dot q^\mu
\dot q^\nu.
\]

The coefficients $\alpha$ and $\beta$ encode the leading
Planck-scale corrections inherited from the curved momentum-space
geometry. In particular, $\alpha$ represents the momentum-space
renormalization of the particle's inertial mass, while $\beta$
captures the leading nonlinear correction to the momentum--velocity
relation arising from the deformed symplectic structure. These
coefficients therefore contain all leading quantum-gravitational
corrections entering the effective matter sector.

To investigate the gravitational consequences of these modified
particle dynamics, we minimally couple the effective action to a
background spacetime metric,
\[
\eta_{\mu\nu}
\rightarrow
g_{\mu\nu}(q),
\]
while leaving the Einstein equations themselves unchanged. Curved
momentum space therefore does not modify the Einstein tensor or the
gravitational field equations directly. Instead, it modifies the
effective matter sector whose stress-energy tensor acts as the source
of the classical Einstein equations. The resulting action becomes

\begin{equation}
S_{\rm eff}^{\rm matter}[q(\tau),g_{\mu\nu}]
=
\int d\tau
\left[
-\alpha
\sqrt{-g_{\mu\nu}(q)\dot q^\mu\dot q^\nu}
-
\beta
\left(
-g_{\mu\nu}(q)\dot q^\mu\dot q^\nu
\right)^{5/2}
\right].
\label{Seff}
\end{equation} 

Since the effective action now plays the role of the matter action in
the semiclassical theory, its variation with respect to the
background metric defines the corresponding Hilbert stress--energy
tensor,
\begin{equation}
T^{\mu\nu}_{\rm eff}(q|q(\tau))
=
-
\frac{2}{\sqrt{-g}}
\frac{\delta S_{\rm eff}^{\rm matter}[q(\tau),g_{\mu\nu}]}
{\delta g_{\mu\nu}(q)}.
\end{equation}

Carrying out the variation gives
\begin{equation}
\begin{aligned}
T^{\mu\nu}_{\rm eff}(q|q(\tau))
=
\int d\tau
\left[
\alpha
(-\dot q^2)^{-1/2}
+
5\beta
(-\dot q^2)^{3/2}
\right]
\frac{\dot q^\mu
\dot q^\nu}
{\sqrt{-g}}
\,
\delta^{(3)}
\!\left(
q-q(\tau)
\right),
\end{aligned}
\label{Teff}
\end{equation}
where
\[
\dot q^2
=
g_{\mu\nu}(q)
\dot q^\mu
\dot q^\nu.
\]

The effective stress-energy tensor provides the central link between
the microscopic momentum-space geometry and the macroscopic spacetime
geometry. The Planck-scale corrections are encoded in the coefficients
$\alpha$ and $\beta$ modify the energy density and momentum flux that
source Einstein's equations, while the gravitational field equations
remain unchanged. Consequently, the deformed BTZ geometry derived in
the following section originates entirely from the modified effective
matter sector rather than from a modification of classical gravity.

In the rest frame of a static particle, the stress tensor reduces to a
localized energy density proportional to $(\alpha+5\beta)$. In the
commutative limit $m_p\rightarrow\infty$, one recovers the standard
stress-energy tensor of a relativistic point particle with mass
$m$, confirming the consistency of the effective semiclassical
construction.

The effective configuration-space action therefore establishes the
semiclassical mechanism through which the microscopic curvature of
momentum space is translated into macroscopic gravitational
phenomena. Curved momentum space modifies the effective matter sector,
whose stress-energy tensor subsequently acts as the source of the
classical Einstein equations, thereby producing the modified BTZ
geometry without altering the Einstein field equations themselves.

\section{Geometry from the Effective Action and Curved Momentum--Space Source}
\label{bg}

We now investigate the classical spacetime geometry sourced by a point
particle whose dynamics are governed by a curved momentum space. The
total action consists of Einstein gravity with negative cosmological
constant $\Lambda_c=-2/\ell^2$, coupled to the effective
configuration--space particle action derived in the previous section,
\begin{equation}
S
=
\frac{1}{16\pi G}
\int d^3x\,\sqrt{-g}\,(R-\Lambda_c)
+
S_{\rm eff}^{\rm matter}[q(\tau),g_{\mu\nu}] .
\end{equation}

Variation of the total action with respect to the spacetime metric
yields Einstein's equations
\begin{equation}
G_{\mu\nu}
+\Lambda_c g_{\mu\nu}
=
8\pi G\,
T_{\mu\nu}^{\mathrm{eff}},
\end{equation}
where the effective stress--energy tensor
\(T_{\mu\nu}^{\mathrm{eff}}\)
is obtained through the Hilbert variation of the effective matter
action and is given by Eq.~(\ref{Teff}).

We now specialize to the static rest-frame configuration of the
particle,
\(
\dot q^\mu=(1,0,0).
\)
Unlike the ordinary relativistic point-particle action, the effective
action derived in the previous section contains higher powers of the
Lorentz invariant quantity
\(
X=-g_{\mu\nu}\dot q^\mu\dot q^\nu
\)
and is therefore not proportional to the proper length of the
worldline. Consequently the evolution parameter \(\tau\) need not
coincide with proper time. Nevertheless, the equations of motion imply
that \(X\) remains constant along the trajectory. We may therefore
normalize the four-velocity such that
\(
\dot q^2=-1,
\)
corresponding to a timelike worldline.

Under this normalization the effective stress--energy tensor becomes
\begin{equation}
T_{\mathrm{eff}}^{\mu\nu}(q)
=
\int d\tau\,
(\alpha+5\beta)
u^\mu u^\nu
\frac{\delta^{(3)}
(q-q(\tau))}
{\sqrt{-g}},
\qquad
u^\mu u_\mu=-1 .
\end{equation}

The above stress tensor is covariantly conserved,
\begin{equation}
\nabla_\mu
T_{\mathrm{eff}}^{\mu\nu}=0,
\end{equation}
as a direct consequence of spacetime diffeomorphism invariance of the
effective action. This conservation law is independent of worldline
reparametrization invariance and guarantees consistency with the
contracted Bianchi identity
\(
\nabla^\mu G_{\mu\nu}=0.
\)

The source is distributional in nature and represents a localized point
particle propagating on the AdS$_3$ background. Taking the trace of the
stress tensor and integrating over time gives the effective spatial
energy density
\begin{equation}
T_{\mathrm{eff}}(q)
=
-Q\,
\delta^{(2)}(\vec q),
\qquad
Q=\alpha+5\beta>0.
\end{equation}
The quantity \(Q\) therefore plays the role of an effective source
strength generated by the curved momentum-space dynamics. Since
\(
Q=M(1+2M^2/m_p^2),
\)
it may be viewed as a second stage of renormalization relating the
effective particle mass \(M\) to the gravitational source entering
Einstein's equations.

Substituting this source into the trace of Einstein's equations yields
the scalar curvature
\begin{equation}
R(q)
=
-\frac{6}{\ell^2}
+
16\pi GQ\,
\delta^{(2)}(\vec q).
\label{scalar}
\end{equation}

It is convenient to separate the curvature into regular and singular
parts,
\begin{equation}
R(q)
=
R_{\mathrm{reg}}
+
R_{\mathrm{sing}}(q),
\end{equation}
where
\begin{equation}
R_{\mathrm{reg}}
=
-\frac{6}{\ell^2},
\qquad
R_{\mathrm{sing}}
=
16\pi GQ\,
\delta^{(2)}(\vec q).
\end{equation}
The regular part describes the constant negative curvature of the
AdS$_3$ background, whereas the singular contribution is localized at
the position of the particle and represents the gravitational
backreaction of the effective source. Away from the origin
(\(r>0\)),
the singular term vanishes, so that the geometry satisfies the vacuum
Einstein equations with a negative cosmological constant and is locally
AdS$_3$.

In $(2+1)$-dimensional gravity a localized source does not generate
propagating gravitational degrees of freedom. Instead, its physical
effect is encoded in the global structure of spacetime through the
conical defect it produces. The relation between the effective source
strength \(Q\), the corresponding deficit angle, and the asymptotic
gravitational mass will now be established. As we shall show, the
curved momentum-space dynamics modify the classical geometry through a
nonlinear relation between the effective source parameter \(Q\) and the
ADM mass measured at the AdS$_3$ boundary. Because the renormalized
particle mass
\(
M=m_p\tan^{-1}(m/m_p)
\)
remains bounded, both the effective source strength \(Q\) and the
resulting ADM mass remain finite. Consequently, the curved
momentum-space structure regularizes the gravitational backreaction
while preserving the classical Einstein dynamics.

Outside the localized source the singular contribution vanishes, and
Einstein's equations reduce to the vacuum equations
\begin{equation}
R_{\mu\nu}
=
-\frac{2}{\ell^{2}}g_{\mu\nu}.
\end{equation}
Consequently, the exterior spacetime is locally AdS$_3$, and the most
general static, circularly symmetric solution is the BTZ family,
\begin{equation}
ds^{2} =
-\left(\frac{r^{2}}{\ell^{2}}-\mu\right) dt^{2}
+ \frac{dr^{2}}
{\left(\frac{r^{2}}{\ell^{2}}-\mu\right)}
+ r^{2} d\phi^{2},
\qquad r>0,
\label{BTZmetric}
\end{equation}
where $\mu$ is a dimensionless parameter characterizing the global
geometry.

When $Q=0$, the metric can be smoothly extended to the origin by taking
$\mu=-1$, corresponding to geodesically complete AdS$_3$ spacetime.
For a non-vanishing effective source, however, the geometry develops a
conical defect whose deficit angle is determined by the parameter
$\mu$. To see this explicitly, consider the spatial section
($t=\mathrm{const}$) of Eq.~(\ref{BTZmetric}),
\begin{equation}
ds^{2}_{(2)}
=
\frac{dr^{2}}
{\left(\frac{r^{2}}{\ell^{2}}-\mu\right)}
+r^{2}d\phi^{2}.
\end{equation}

Near the origin $(r\approx0)$, for $\mu<0$, introduce the radial
coordinate
\[
\rho=\frac{r}{\sqrt{-\mu}},
\]
so that
\begin{equation}
ds^{2}_{(2)}
\simeq
d\rho^{2}
+(-\mu)\rho^{2}d\phi^{2}.
\end{equation}
The total angle around the origin is therefore
\[
\delta\phi
=
2\pi\sqrt{-\mu},
\]
and the corresponding deficit angle becomes
\begin{equation}
\Delta\phi
=
2\pi-\delta\phi
=
2\pi
\left(
1-\sqrt{-\mu}
\right).
\label{deficit}
\end{equation}

The same relation may also be obtained from the holonomy of the spin
connection, as shown in Appendix~F.

We now relate the effective source strength $Q$ to the deficit angle
using the Gauss--Bonnet theorem. Integrating the singular curvature over
an infinitesimal disc $D$ surrounding the origin gives
\begin{equation}
\int_D
R_{\rm sing}
\sqrt{g^{(2)}}\,d^{2}x
=
2\Delta\phi.
\label{GB}
\end{equation}
Using
\begin{equation}
R_{\rm sing}
=
16\pi GQ
\,
\frac{\delta^{(2)}(\tilde{x})}
{\sqrt{g^{(2)}}},
\end{equation}
one immediately finds
\begin{equation}
\Delta\phi
=
8\pi GQ.
\label{deltaQ}
\end{equation}

Combining Eqs.~(\ref{deficit}) and (\ref{deltaQ}) yields the relation
between the effective source strength and the BTZ parameter,
\begin{equation}
Q
=
\frac{1}{4G}
\left(
1-\sqrt{-\mu}
\right),
\label{Qmu}
\end{equation}
or equivalently,
\begin{equation}
\mu
=
-(1-4GQ)^{2}.
\label{muQ1}
\end{equation}

Writing Eq.~(\ref{muQ1}) in the symmetric form
\begin{equation}
\sqrt{|\mu|}
=
|1-4GQ|
\end{equation}
makes the analytic structure of the solution manifest. The absolute-value
representation extends continuously across the critical value
\[
Q_c=\frac{1}{4G},
\]
and separates the geometry into two distinct branches,
\begin{equation}
\mu(Q)=
\begin{cases}
-(1-4GQ)^2, & Q<\dfrac{1}{4G},\\[2mm]
0, & Q=\dfrac{1}{4G},\\[2mm]
(4GQ-1)^2, & Q>\dfrac{1}{4G}.
\end{cases}
\end{equation}

The first branch corresponds to locally AdS$_3$ spacetimes containing a
conical defect, whereas the second branch describes BTZ black holes.
The transition occurs at the critical value
\begin{equation}
Q_c=\frac{1}{4G},
\end{equation}
for which the deficit angle reaches
\[
\Delta\phi=2\pi,
\]
the conical geometry degenerates, and the solution becomes the massless
BTZ spacetime with $\mu=0$. Beyond this point the geometry enters the
massive BTZ branch with $\mu>0$.

The parameter $\mu$ determines the local geometry outside the source.
The physical mass of the spacetime, however, is not identified with
$\mu$ directly but is defined through the conserved charges associated
with the asymptotic AdS$_3$ boundary conditions. In asymptotically
AdS$_3$ spacetimes, the conserved energy is obtained from the
Brown--York quasi-local stress tensor~\cite{PhysRevD.47.1407}, which is
equivalent to the Brown--Henneaux asymptotic charge construction
\cite{BrownHenneaux1986}.

The induced metric on a timelike hypersurface $r=\mathrm{const}$ is
\begin{equation}
h_{ab}dx^adx^b
=
-f^2(r)\,dt^2+r^2d\phi^2,
\qquad
f^2(r)=\frac{r^2}{\ell^2}-\mu.
\end{equation}

The renormalized Brown--York stress tensor is
\begin{equation}
T_{ab}
=
\frac{1}{8\pi G}
\left(
K_{ab}
-
Kh_{ab}
+
\frac{1}{\ell}h_{ab}
\right),
\end{equation}
where $K_{ab}$ denotes the extrinsic curvature of the boundary and the
last term is the standard counterterm required for asymptotically
AdS$_3$ geometries.

The ADM mass is the conserved charge associated with the timelike
Killing vector $\partial_t$,
\begin{equation}
M_{\rm ADM}
=
\oint
\sqrt{\sigma}\,
u^a
(\partial_t)^b
T_{ab}\,
d\phi,
\end{equation}
where $u^a$ is the future-directed unit normal to the spatial slice of
the boundary.

Evaluating the above expression in the limit
$r\rightarrow\infty$ gives
\begin{equation}
M_{\rm ADM}
=
\frac{\mu}{8G}.
\label{ADMmass}
\end{equation}

Equation~(\ref{ADMmass}) shows that the asymptotic gravitational mass is
completely determined by the parameter $\mu$, which itself is fixed by
the effective source strength through Eq.~(\ref{muQ1}). Consequently,
the ADM mass is a nonlinear function of the effective source,
\begin{equation}
M_{\rm ADM}
=
\frac{1}{8G}
(4GQ-1)^2,
\qquad
Q>\frac{1}{4G},
\end{equation}
for the BTZ branch, while the particle branch corresponds to
\begin{equation}
\mu
=
-(1-4GQ)^2,
\qquad
Q<\frac{1}{4G}.
\end{equation}

It is important to emphasize that the effective parameter $Q$ is not
itself the gravitational mass. Rather, it characterizes the localized
matter source generated by the curved momentum-space dynamics. The
physical mass measured at the AdS$_3$ boundary is the conserved
Brown--York (equivalently Brown--Henneaux) charge, which includes the
effects of gravitational binding energy through the nonlinear relation
between $Q$ and $\mu$.

The Gauss--Bonnet construction also admits an equivalent topological
interpretation. In $(2+1)$-dimensional gravity, the absence of local
propagating degrees of freedom implies that the geometry generated by a
localized source is completely characterized by the holonomy of the
connection around the defect. Consequently, the conical defect obtained
from the Gauss--Bonnet theorem admits an equivalent description in terms
of the AdS-subtracted holonomy.

To illustrate this correspondence, consider the two-dimensional spatial
slice of the spacetime. The torsion-free spin connection satisfies
Cartan's second structure equation,
\begin{equation}
d\omega=\frac{1}{2}R\sqrt{g}\,d^2x,
\end{equation}
which, upon applying Stokes' theorem, yields
\begin{equation}
2\oint_{\partial\Sigma}\omega
=
\int_{\Sigma}R\sqrt{g}\,d^2x.
\end{equation}
Subtracting the corresponding AdS$_3$ contribution defines the
AdS-subtracted holonomy,
\begin{equation}
H
=
2\left(
\oint_{\partial\Sigma}\omega
-
\oint_{\partial\Sigma}\omega_{\rm AdS}
\right)
=
\int_{\Sigma}
\sqrt{g}
\left(
R+\frac{6}{\ell^{2}}
\right)d^{2}x.
\end{equation}
Using Eq.~(\ref{scalar}), we immediately obtain
\begin{equation}
H
=
16\pi GQ.
\end{equation}
Since
\[
R+\frac{6}{\ell^{2}}=0
\]
everywhere outside the localized source, the value of $H$ is independent
of the integration contour. The AdS-subtracted holonomy is therefore a
topological invariant determined entirely by the localized source and
contains precisely the same information as the integrated curvature or,
equivalently, the conical deficit angle obtained from the Gauss--Bonnet
theorem.

The Cartan-geometric argument presented above serves only as a geometric
illustration of this correspondence. A complete derivation in the
first-order Chern--Simons formulation of $(2+1)$-dimensional gravity is
given in Appendix~F of Ref.~\cite{Nandi:2025mco}, where the Wilson loop
of the Chern--Simons connection is evaluated explicitly and shown to
reproduce the same relation between the holonomy and the deficit angle.

The overall logical structure of the construction may therefore be
summarized schematically as
\begin{equation}
m
\;\longrightarrow\;
M
\;\longrightarrow\;
Q
\;\xrightarrow{\mathrm{Einstein\ equations}}\;
R_{\rm sing}
\;\xrightarrow{\mathrm{Geometry}}\;
\left\{
\begin{array}{c}
\Delta\phi\\[1mm]
H
\end{array}
\right.
\;\xrightarrow{}\;
\mu(Q)
\;\xrightarrow{\mathrm{Brown\!-\!York/Brown\!-\!Henneaux}}\;
M_{\rm ADM}.
\end{equation}

This sequence clearly illustrates how the microscopic curved
momentum-space dynamics are transmitted to the classical gravitational
sector. The bare particle mass $m$ is first renormalized to the
effective particle mass
\(
M=m_p\tan^{-1}(m/m_p),
\)
which subsequently determines the effective source strength
\(
Q=M(1+2M^2/m_p^2)
\)
appearing in Einstein's equations. The localized curvature generated by
this source admits two equivalent descriptions: a geometric one through
the conical deficit angle $\Delta\phi$ and a topological one through the
AdS-subtracted holonomy $H$. Both determine the BTZ parameter
$\mu(Q)$, from which the conserved Brown--York (equivalently
Brown--Henneaux) ADM mass is obtained. Since the renormalized mass
$M$ remains bounded even for arbitrarily large bare mass, the effective
source strength $Q$ and hence the corresponding ADM mass also remain
finite. The curved momentum-space structure therefore regularizes the
gravitational backreaction while the exterior spacetime remains locally
AdS$_3$.

\subsection{Deformed BTZ Geometry and Thermodynamics}
\label{subsec:BTZ_thermo}

In $(2+1)$-dimensional gravity with a negative cosmological constant,
$\Lambda_c<0$, a point source gives rise to a spacetime that is locally
AdS$_3$, while its global properties are completely determined by the
holonomy around the source. Consequently, the corresponding geometry can
be written in the standard BTZ form,
\begin{equation}
ds^2
=
-f^2(r)\,dt^2
+\frac{dr^2}{f^2(r)}
+r^2d\phi^2.
\label{metric}
\end{equation}

In the present framework, however, the geometry is determined by a
nonlinear relation between the microscopic mass parameter $M$ and the
physical ADM mass. Introducing
\begin{equation}
Q(M)
=
\alpha+5\beta
=
M\left(1+\frac{2M^2}{m_p^2}\right),
\label{v}
\end{equation}
together with
\begin{equation}
A_{m_p}(M)
=
4GQ(M)-1,
\label{w}
\end{equation}
the ADM mass obtained in Sec.~\ref{bg} is
\begin{equation}
M_{\rm ADM}
=
\frac{A_{m_p}^2(M)}{8G}.
\label{mn}
\end{equation}

It is useful to relate this parametrization to the conventional BTZ
description. In the standard solution, the geometry is characterized by
the parameter $\mu$, which satisfies
\begin{equation}
\mu=(4GQ-1)^2.
\end{equation}
Within the present construction this becomes simply
\begin{equation}
\mu=A_{m_p}^2(M),
\end{equation}
demonstrating that the function $A_{m_p}(M)$ provides a natural
generalization of the BTZ mass parameter in the presence of curved
momentum-space effects.

Substituting Eq.~(\ref{mn}) into the BTZ lapse function gives
\begin{equation}
f^2(r)
=
-8GM_{\rm ADM}
+\frac{r^2}{\ell^2}
=
-A_{m_p}^2(M)
+\frac{r^2}{\ell^2}.
\end{equation}

The event horizon is determined by the largest positive root of
$f^2(r)=0$. In the black-hole regime,
\begin{equation}
A_{m_p}(M)>0,
\end{equation}
the horizon radius is therefore
\begin{equation}
r_+
=
\ell\,A_{m_p}(M)
=
\ell\sqrt{8GM_{\rm ADM}}.
\end{equation}

The surface gravity is given by
\begin{equation}
\kappa
=
\frac12
\left.
\frac{df^2(r)}{dr}
\right|_{r=r_+}
=
\frac{r_+}{\ell^2},
\end{equation}
which immediately yields the Hawking temperature
\begin{equation}
T_H
=
\frac{\kappa}{2\pi}
=
\frac{r_+}{2\pi\ell^2}
=
\frac{A_{m_p}(M)}{2\pi\ell}.
\end{equation}

Assuming the validity of the Bekenstein--Hawking entropy formula in
$(2+1)$ dimensions~\cite{carlip1995b}, the entropy (in units
$\hbar=k_B=1$) becomes
\begin{equation}
S
=
\frac{2\pi r_+}{4G}
=
\frac{\pi\ell}{2G}\,
A_{m_p}(M)
=
\frac{\pi\ell}{2G}
\sqrt{8GM_{\rm ADM}}.
\end{equation}

We emphasize that the appropriate thermodynamic energy is the ADM mass,
Eq.~(\ref{mn}), which is determined by the global geometry—or,
equivalently, by the holonomy class (see Appendix~F)—and represents the
total conserved energy measured by an observer at the asymptotic
boundary. Although the effective description introduces a nonlinear
relation between the microscopic parameter $M$ and the geometry, the
thermodynamic quantities depend exclusively on the ADM mass
$M_{\rm ADM}$, which fully incorporates the gravitational backreaction.

Using the standard BTZ relations,
\begin{equation}
T_H
=
\frac{r_+}{2\pi\ell^2},
\qquad
S
=
\frac{2\pi r_+}{4G},
\end{equation}
together with
\begin{equation}
r_+
=
\ell\sqrt{8GM_{\rm ADM}},
\end{equation}
the first law of black-hole thermodynamics follows immediately.
Differentiating,
\begin{equation}
dS
=
\frac{2\pi}{4G}\,dr_+,
\qquad
dM_{\rm ADM}
=
\frac{r_+}{4G\ell^2}\,dr_+,
\end{equation}
one obtains
\begin{equation}
dM_{\rm ADM}
=
T_H\,dS,
\end{equation}
demonstrating that the thermodynamic structure remains identical to that
of the classical BTZ black hole.

Thus, all Planck-scale corrections are encoded entirely in the nonlinear
mapping between the microscopic mass parameter $M$ and the ADM mass
$M_{\rm ADM}$. The local BTZ geometry and its thermodynamic relations
remain unchanged, while the deformation modifies the global conserved
charges through the effective momentum-space geometry.

\subsubsection*{Hamilton--Jacobi Tunneling Method}
\label{subsubsec:tunneling_general}

Hawking radiation may be interpreted as a quantum tunneling process in which particles escape across the event horizon. Within the semiclassical Hamilton--Jacobi formulation, the tunneling probability is determined by the imaginary part of the classical action acquired when a particle traverses the horizon. Since this derivation depends only on the near-horizon structure of the metric, it is applicable to both the classical BTZ geometry and its deformed counterpart considered in this work.

For a static, circularly symmetric spacetime described by
\begin{equation}
ds^{2}
=
-f^{2}(r)\,dt^{2}
+\frac{dr^{2}}{f^{2}(r)}
+r^{2}d\phi^{2},
\end{equation}
the lapse function possesses a simple zero at the event horizon,
\begin{equation}
f^{2}(r)
\simeq
\kappa (r-r_{+}),
\end{equation}
where $\kappa$ denotes the surface gravity.

The Hamilton--Jacobi equation for a massless particle is
\begin{equation}
g^{\mu\nu}\partial_{\mu}I\,\partial_{\nu}I=0,
\end{equation}
where $I$ is the classical action. Restricting attention to radial motion and adopting the standard ansatz
\begin{equation}
I=-\omega t+W(r),
\end{equation}
one obtains
\begin{equation}
-f^{-2}(r)\omega^{2}
+
f^{2}(r)\left(W'(r)\right)^{2}
=0,
\end{equation}
which immediately gives
\begin{equation}
W'(r)
=
\frac{\omega}{f^{2}(r)}.
\end{equation}

The radial action therefore takes the form
\begin{equation}
W(r)
=
\int
\frac{\omega}{f^{2}(r)}\,dr.
\end{equation}
Near the horizon the integrand develops a simple pole. Evaluating the contour integral using the standard Feynman prescription yields
\begin{equation}
\mathrm{Im}\,I
=
\mathrm{Im}\,W
=
\frac{\pi\omega}{\kappa}.
\end{equation}
Consequently, the tunneling probability is
\begin{equation}
\Gamma
\sim
e^{-2\,\mathrm{Im}I}
=
\exp\!\left(-\frac{2\pi\omega}{\kappa}\right)
=
\exp\!\left(-\frac{\omega}{T_H}\right),
\end{equation}
where
\begin{equation}
T_H
=
\frac{\kappa}{2\pi}
\end{equation}
is the Hawking temperature.

A crucial refinement introduced by Parikh and Wilczek is the incorporation of energy conservation. During the emission of a quantum with energy $\omega$, the black hole loses the same amount of energy, causing the horizon to shrink continuously throughout the tunneling process. The imaginary part of the action is therefore computed as
\begin{equation}
\mathrm{Im}\,I
=
\int_{r_{\rm in}}^{r_{\rm out}}
\int_{0}^{\omega}
\frac{d\omega'}
{f^{2}(r;M-\omega')}
\,dr,
\end{equation}
where the geometry is evaluated using the instantaneous black-hole mass.

Remarkably, the resulting emission probability can be expressed solely in terms of the change in the Bekenstein--Hawking entropy,
\begin{equation}
\Gamma(\omega)
\sim
e^{\Delta S},
\qquad
\Delta S
=
S(M-\omega)-S(M),
\end{equation}
demonstrating that backreaction renders the Hawking spectrum slightly nonthermal and allows for correlations between successive emissions.

In the following subsection we apply this general formalism to the deformed BTZ black hole. In that case, the tunneling calculation retains exactly the same structure; the only modification is that the geometry is determined by the nonlinear relation between the microscopic mass parameter and the ADM mass derived in Sec.~\ref{bg}.

\subsubsection*{Application to the Deformed BTZ Black Hole}
\label{subsubsec:tunneling_BTZ}

We now apply the Hamilton--Jacobi tunneling formalism to the deformed BTZ black hole discussed in Sec.~\ref{bg}. Owing to the curvature of momentum space, the spacetime geometry is not determined directly by the microscopic mass parameter $M$, but rather by the corresponding ADM mass,
\begin{equation}
M_{\rm ADM}(M)
=
\frac{1}{8G}
\left[
4GM\!\left(1+\frac{2M^2}{m_p^2}\right)-1
\right]^2
=
\frac{A_{m_p}^2(M)}{8G},
\label{f}
\end{equation}
where
\begin{equation}
A_{m_p}(M)
=
4GM\!\left(1+\frac{2M^2}{m_p^2}\right)-1 .
\end{equation}

The corresponding Bekenstein--Hawking entropy is therefore
\begin{equation}
S(M_{\rm ADM})
=
\frac{\pi\ell}{2G}
\sqrt{8GM_{\rm ADM}}
=
\frac{\pi\ell}{2G}
A_{m_p}(M).
\end{equation}

Within the tunneling picture, the emission of a Hawking quantum with energy $\omega$ is accompanied by a decrease of the ADM mass,
\begin{equation}
M_{\rm ADM}
\longrightarrow
M_{\rm ADM}-\omega .
\end{equation}
If the black-hole state is parametrized by the microscopic mass parameter $M$, the corresponding value $M'$ after emission is determined implicitly through
\begin{equation}
M_{\rm ADM}(M')
=
M_{\rm ADM}(M)-\omega .
\label{hg}
\end{equation}
Expanding to first order in $\omega$ gives
\begin{equation}
M'
\simeq
M
-
\frac{\omega}
{
A_{m_p}(M)
\left(
1+\frac{6M^2}{m_p^2}
\right)
}.
\end{equation}

The resulting entropy change is
\begin{equation}
\Delta S
=
S(M_{\rm ADM}-\omega)-S(M_{\rm ADM})
\simeq
-\omega
\frac{dS}{dM_{\rm ADM}},
\end{equation}
where
\begin{equation}
\frac{dS}{dM_{\rm ADM}}
=
\frac{2\pi\ell}
{\sqrt{8GM_{\rm ADM}}}.
\end{equation}
Consequently, the tunneling probability becomes
\begin{equation}
\Gamma(\omega)
\sim
\exp\!\left[
-\omega
\frac{2\pi\ell}
{\sqrt{8GM_{\rm ADM}}}
\right]
=
\exp\!\left(
-\frac{\omega}{T_H}
\right),
\end{equation}
with the Hawking temperature
\begin{equation}
T_H(M) 
=
\frac{1}{2\pi \ell}
\left[
4GM\!\left(1+\frac{2M^2}{m_p^2}\right)-1
\right].
\end{equation}

This result is in complete agreement with the Hawking temperature obtained from the surface gravity in Sec.~\ref{bg}. Therefore, the Parikh--Wilczek tunneling formalism remains unchanged in the presence of curved momentum space; the Planck-scale modification is encoded entirely in the nonlinear relation between the microscopic mass parameter $M$ and the ADM mass that sources the geometry.

\subsection{Return Time of an Emitted Hawking Quantum with Backreaction on the BTZ Geometry}

We now reinterpret the massless particle propagating between the event horizon and the $\mathrm{AdS}_{3}$ boundary as an emitted Hawking quantum of energy $\omega$, rather than as a passive probe considered in the previous subsection. In this case, energy conservation requires that the emission dynamically backreacts on the spacetime geometry through a reduction of the ADM mass,
\begin{equation}
M_{\rm ADM}(M)
\longrightarrow
M_{\rm ADM}(M)-\omega,
\end{equation}
where $M_{\rm ADM}(M)$ is given by Eq.~(\ref{f}).

The corresponding lapse function becomes
\begin{equation}
f^{2}(r;\omega)
=
\frac{r^{2}}{\ell^{2}}
-
8G
\left[
M_{\rm ADM}(M)-\omega
\right].
\end{equation}

Since
\begin{equation}
M_{\rm ADM}(M)
=
\frac{A_{m_p}^{2}(M)}{8G},
\end{equation}
the horizon radius after emission is
\begin{equation}
r_{+}(\omega)
=
\ell
\sqrt{
A_{m_p}^{2}(M)-8G\omega
},
\end{equation}
where
\begin{equation}
A_{m_p}(M)
=
4GM\left(1+\frac{2M^{2}}{m_{p}^{2}}\right)-1.
\end{equation}

Proceeding exactly as in the probe analysis, the return time of the emitted Hawking quantum is
\begin{equation}
\tau_{\rm ret}(\omega)
=
2
\int_{r_{0}}^{\infty}
\frac{dr}
{\dfrac{r^{2}}{\ell^{2}}
-
8G
\left(M_{\rm ADM}-\omega\right)}
=
\frac{\ell^{2}}
{r_{+}(\omega)}
\ln
\left(
\frac{r_{0}+r_{+}(\omega)}
{r_{0}-r_{+}(\omega)}
\right).
\end{equation}

Choosing the emission point as
\begin{equation}
r_{0}
=
r_{+}(\omega)(1+s),
\qquad
s>0,
\end{equation}
the near-horizon expression becomes
\begin{equation}
\tau_{\rm ret}(\omega)
=
\frac{\ell^{2}}
{r_{+}(\omega)}
\ln\!\left(1+\frac{2}{s}\right).
\end{equation}

For $\omega\ll M_{\rm ADM}$, the horizon radius may be expanded as
\begin{equation}
r_{+}(\omega)
=
\ell A_{m_p}(M)
\left[
1-
\frac{4G\omega}
{A_{m_p}^{2}(M)}
+
\mathcal O(\omega^{2})
\right].
\end{equation}

Substituting this expression into the return time gives
\begin{equation}
\tau_{\rm ret}(\omega)
\simeq
\frac{\ell}
{A_{m_p}(M)}
\left[
1+
\frac{4G\omega}
{A_{m_p}^{2}(M)}
\right]
\ln\!\left(1+\frac{2}{s}\right),
\end{equation}
or equivalently,
\begin{equation}
\tau_{\rm ret}(\omega)
\simeq
\tau_{\rm ret}(M)
\left[
1+
\frac{4G\omega}
{A_{m_p}^{2}(M)}
\right].
\end{equation}

At this point it is useful to distinguish two physically different sources of corrections to the return time.

First, curved momentum space modifies the background geometry through the nonlinear relation between the microscopic mass parameter and the ADM mass. As shown in the previous subsection, this gives
\begin{equation}
\frac{\tau_{\rm ret}^{\rm(mod)}}
{\tau_{\rm ret}^{\rm(class)}}
=
\epsilon,
\end{equation}
where
\begin{equation}
\epsilon
=
\frac{4Gm-1}
{4GM\!\left(1+\dfrac{2M^{2}}{m_{p}^{2}}\right)-1}.
\end{equation}

Expanding in the perturbative regime $m^{2}/m_{p}^{2}\ll1$,
\begin{equation}
\epsilon
\simeq
1-
\frac{4Gm^{3}}
{m_{p}^{2}(4Gm-1)}
+
\mathcal O\!\left(\frac{m^{4}}{m_{p}^{4}}\right).
\end{equation}

Second, Hawking emission produces an additional dynamical correction,
\begin{equation}
\frac{
\tau_{\rm ret}(\omega)
-
\tau_{\rm ret}(M)}
{\tau_{\rm ret}(M)}
\simeq
\frac{4G\omega}
{A_{m_p}^{2}(M)}.
\end{equation}

Combining both effects, the total fractional correction to the return time is
\begin{equation}
\frac{\tau_{\rm ret}^{\rm mod}(\omega)-\tau_{\rm ret}^{\rm class}}
{\tau_{\rm ret}^{\rm class}}
\simeq
-\underbrace{\frac{4Gm^{3}}
{m_{p}^{2}(4Gm-1)}}_{\text{geometric correction}}
+
\underbrace{\frac{4G\omega}
{4Gm-1}}_{\text{Hawking backreaction}}
+\mathcal{O}\!\left(\frac{\omega}{m_{p}^{2}}\right).
\label{eq:total_correction}
\end{equation}

Equation~(\ref{eq:total_correction}) clearly separates two physically distinct contributions to the return time. The first term is a purely geometric correction arising from the deformation of the background spacetime induced by curved momentum space and is independent of the energy of the emitted Hawking quantum. The second term originates from the dynamical backreaction associated with Hawking radiation and depends explicitly on the emitted energy $\omega$. These contributions act in opposite directions: the geometric deformation decreases the return time by modifying the effective horizon geometry, whereas Hawking backreaction increases it through the reduction of the black-hole mass during the emission process.

\begin{figure}[t]
\centering
\includegraphics[width=0.72\linewidth]{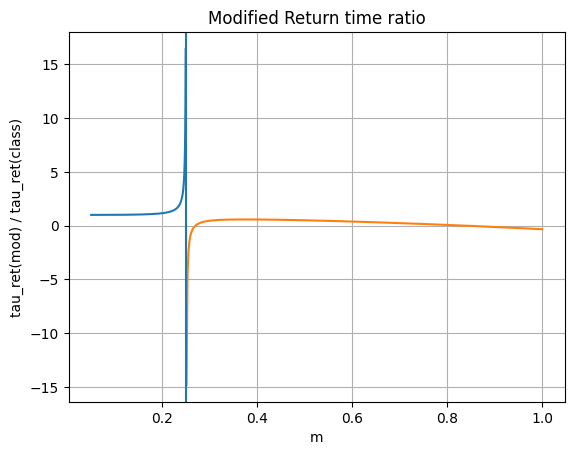}
\caption{
Ratio of the modified return time to the classical return time as a function of the mass parameter $m$. The blue branch ($m<1/4G$) corresponds to the conical-defect (particle) sector, whereas the orange branch ($m>1/4G$) represents the BTZ black-hole sector. The apparent pole at $m=1/(4G)$ reflects the perturbative enhancement of the return-time correction near the geometric transition. In the exact theory, the transition is determined by $A_{m_p}(M)=0$, and the perturbative pole provides only an approximate indication of the critical point. The parameters are chosen as $G=1$ and $m_p=1$.
}
\label{fig:tau_ratio}
\end{figure}

The correction exhibits a pole at $4Gm=1$, signalling a pronounced enhancement of the return-time modification as the system approaches the transition between the conical-defect and BTZ phases. This behaviour is illustrated in Fig.~\ref{fig:tau_ratio}, which shows the ratio of the modified return time to its classical counterpart as a function of the mass parameter $m$. The rapid increase near $4Gm=1$ reflects the growing sensitivity of the return time to small changes in the background geometry close to the critical regime.

It is important to stress that this pole is not a physical singularity of the spacetime. Rather, it arises from the perturbative expansion expressed in terms of the classical mass parameter $m$. In the exact theory, the geometric transition is determined by
\begin{equation}
A_{m_p}(M)=0,
\qquad\text{or equivalently}\qquad
4GM\left(1+\frac{2M^{2}}{m_{p}^{2}}\right)=1,
\end{equation}
which reduces to the classical condition $4GM=1$ in the limit $m_p\rightarrow\infty$. Since the perturbative parameter $m$ differs from the microscopic mass parameter $M$ by Planck-scale corrections, the apparent pole at $4Gm=1$ should be regarded as an approximate signature of the nearby geometric transition rather than its exact location.

More generally, these results demonstrate that the modification of the return time originates from changes in the effective spacetime geometry rather than from any alteration of the local propagation law. The null geodesic equations retain their classical form; instead, the nonlinear relation between the microscopic mass parameter and the ADM mass modifies the horizon radius and consequently the geometry through which the signal propagates.

Although proper time is not defined for null trajectories, their evolution may be parametrized by an affine parameter. As shown in Appendix~D, this affine parameter diverges as the horizon is approached, reflecting the infinite redshift associated with the BTZ horizon. In the next subsection we extend the analysis beyond the fixed-background approximation by incorporating Hawking radiation, thereby allowing the ADM mass to evolve dynamically during the emission process.

In the classical limit, $m_p\rightarrow\infty$, the momentum-space deformation disappears and the geometric contribution vanishes, leaving only the standard Hawking backreaction. Once Planck-scale effects are included, however, the geometric correction becomes the dominant contribution near the critical regime, while the Hawking-induced correction remains parametrically suppressed by the small ratio $\omega/m$ characteristic of a typical Hawking quantum. Consequently, the return-time observable is primarily controlled by the underlying Planck-scale deformation of the geometry, with Hawking backreaction providing a comparatively small perturbative correction.

\section{Conclusion}

One of the central challenges of quantum gravity is that the microscopic structure of spacetime remains unknown. Existing approaches, including spin-foam models, string theory, causal dynamical triangulations, causal sets, group field theory, and other frameworks, propose different microscopic degrees of freedom, yet no experimental evidence presently identifies the correct description of Planck-scale physics. Consequently, while these approaches differ significantly at the microscopic level, it is natural to ask whether they admit common low-energy manifestations that can be investigated independently of their underlying realization.

An important example of such an effective description was provided by Freidel and Livine~\cite{Freidel:2005bb,PhysRevLett.96.221301}, who demonstrated that, within the Ponzano--Regge spin-foam model of $(2+1)$-dimensional quantum gravity, integrating over the microscopic gravitational degrees of freedom naturally leads to an effective noncommutative field theory with a curved (group-valued) momentum space. Their work establishes curved momentum space as an emergent low-energy consequence of a specific microscopic quantum gravity model.

The present work adopts a complementary perspective. Rather than assuming a
specific microscopic realization of quantum gravity, we begin directly with
the effective geometry of momentum space. Our motivation is that,
irrespective of the underlying theory of quantum gravity, any
experimentally accessible signature must ultimately emerge through its
low-energy effective manifestation. We therefore regard curved momentum
space not as a fundamental description of quantum gravity, but as an
effective characterization of a particular quantum corner of the
Planck-scale structure of spacetime. By a quantum corner we mean a particular effective sector of an
otherwise unknown UV-complete theory of quantum gravity, in which the
relevant low-energy quantum kinematics can be described by an algebra of
spacetime localization observables. Within the present
framework, this quantum corner is modeled by a Lie-algebraic
noncommutative geometry, specifically the $\mathrm{su}(1,1)$ algebra of
spacetime localization operators. The semiclassical limit of this
localization algebra naturally reconstructs an effective curved
momentum-space geometry. Our objective is therefore not to identify the
microscopic constituents of spacetime, but rather to understand how this
effective momentum-space geometry manifests itself in classical
gravitational phenomena.

The present work adopts a complementary perspective. Rather than assuming a specific microscopic realization of quantum gravity, we begin directly with the effective geometry of momentum space. Our motivation is that, irrespective of the underlying theory of quantum gravity, any experimentally accessible signature must ultimately emerge through its low-energy effective manifestation. We therefore regard curved momentum space not as a fundamental description of quantum gravity, but as an effective characterization of a particular quantum corner of the Planck-scale structure of spacetime. At the microscopic level, this quantum corner is described by a Lie-algebraic noncommutative geometry, specifically one governed by the $\mathrm{su}(1,1)$ algebra of spacetime localization operators. The semiclassical limit of this algebraic structure naturally reconstructs an effective curved momentum-space geometry. Our objective is therefore not to identify the microscopic constituents of spacetime, but rather to understand how this emergent momentum-space geometry manifests itself in classical gravitational phenomena.

Within this framework, spacetime geometry is reconstructed rather than postulated. Starting from the deformed phase-space algebra, we show that the semiclassical limit naturally reconstructs a locally $\mathrm{AdS}_3$ momentum-space geometry, whose geodesic distance defines the physical particle mass. The resulting momentum-space geometry deforms the symplectic structure while preserving Lorentz symmetry and naturally yields a finite renormalized particle mass. Consequently, it provides an intrinsic ultraviolet regularization of point-particle sources without modifying the underlying classical spacetime manifold.

To establish the gravitational consequences of these microscopic kinematical effects, we constructed an effective configuration-space action using Darboux variables appropriate for the semiclassical regime. Although the effective spacetime coordinates become commutative, the information encoded in the underlying noncommutative algebra survives through momentum-dependent vielbeins and higher-order corrections. From this action we derived the corresponding effective energy-momentum tensor and demonstrated explicitly that it consistently sources the ordinary Einstein equations. Consequently, quantum-gravitational information encoded in momentum-space geometry is dynamically transferred into classical spacetime through the standard Einstein equations, without requiring any modification of the gravitational field equations themselves.

Applying this framework to Einstein gravity with a negative cosmological constant, we obtained a deformed BTZ black-hole solution whose ADM mass, horizon radius, Hawking temperature, and Bekenstein--Hawking entropy are determined through a nonlinear relation between the microscopic particle mass and the conserved gravitational mass. Although the functional form of the BTZ thermodynamic relations remains unchanged, the conserved quantities receive finite Planck-scale corrections inherited entirely from the momentum-space geometry. We further investigated Hawking radiation within the Hamilton--Jacobi tunneling formalism and showed that the tunneling framework itself remains unchanged. Instead, all quantum-gravity corrections are encoded in the modified spacetime geometry generated through the effective matter sector. The return time of Hawking quanta likewise receives both geometric and Hawking-backreaction contributions, with the geometric correction dominating near the transition between the conical-defect and BTZ sectors. An important conceptual consequence is that the null geodesic equations remain unchanged; the modified propagation time arises solely from the deformation of the effective spacetime geometry.

The principal contribution of the present work is therefore not the introduction of curved momentum space itself. Previous studies, including those of Freidel and Livine~\cite{Freidel:2005bb,PhysRevLett.96.221301}, have shown how curved momentum space emerges from specific microscopic quantum gravity models and have explored its implications for noncommutative geometry, modified particle kinematics, deformed dispersion relations, and relative locality. However, these investigations primarily treat momentum-space geometry as an effective kinematical framework describing particle propagation on a prescribed background. The question of how such an effective momentum-space geometry feeds back into the gravitational field and modifies the spacetime itself has remained largely unexplored.

The central result of this work is the explicit demonstration of this missing link. Taking curved momentum space as the effective low-energy manifestation of a particular quantum corner of an otherwise unknown UV-complete theory of quantum gravity, we have shown that it generates a well-defined effective energy-momentum tensor that consistently sources the classical Einstein equations. This establishes a complete semiclassical mechanism through which the effective quantum kinematics encoded in momentum-space geometry backreacts on macroscopic spacetime geometry. Within this framework, curved momentum space is elevated from a purely kinematical concept to a dynamical intermediary that transfers the effective quantum imprint of the Planck-scale structure of spacetime into classical spacetime through the unmodified Einstein equations. Consequently, quantum-gravity effects become manifest not through modifications of Einstein's theory of gravity itself, but through the effective matter sector generated by the underlying momentum-space geometry, which dynamically sources the classical Einstein equations.

More broadly, our results suggest a complementary strategy for investigating the semiclassical regime of quantum gravity. Instead of attempting to identify the microscopic constituents of spacetime, which remain theoretically model-dependent and experimentally inaccessible, one may investigate their universal low-energy manifestations. If curved momentum space indeed represents such an effective manifestation of Planck-scale physics, then many observable quantum-gravity effects can be studied independently of the underlying microscopic realization. In this picture, momentum-space geometry encodes the effective quantum imprint of spacetime, while gravity acts as the dynamical bridge that converts microscopic quantum kinematics into macroscopic spacetime geometry through the unmodified Einstein equations.

Although the present analysis has been carried out in $(2+1)$ dimensions, the mechanism developed here is sufficiently general to motivate extensions to spinning particles, interacting systems, higher-dimensional gravity, and cosmological spacetimes. An important direction for future work is to determine whether the same momentum-space-induced backreaction mechanism persists in $(3+1)$ dimensions. If so, the framework developed here may provide a general semiclassical paradigm for investigating observable quantum-gravity effects without requiring prior knowledge of the fundamental microscopic structure of spacetime.

\section*{Acknowledgements}

I am grateful to the organizers of the South African Gravity Society (SAGS) 2025 Conference for their kind invitation and for organizing a stimulating and enjoyable meeting. I also thank my colleagues and collaborators for many valuable discussions on momentum-space geometry and $(2+1)$-dimensional gravity. This work was supported by Stellenbosch University and the National Institute for Theoretical and Computational Sciences (NITheCS), South Africa. I am particularly grateful to Salvatore Mignemi (University of Cagliari) for his kind invitation, warm hospitality, and many insightful discussions during my research visit, during which the final part of this work was completed. I would also like to thank Pei-Ming Ho  and George Zoupanos for their useful correspondence. Finally, I sincerely thank M. Roy, L. Horoto, B. Chakraborty, and F. G. Scholtz for many insightful discussions, fruitful collaborations, and valuable contributions to the development of this work.



\bibliographystyle{unsrt}
\bibliography{bibtexfile1}

\end{document}